\documentclass[conference]{IEEEtran}
\IEEEoverridecommandlockouts
\usepackage{cite}
\usepackage{amsmath,amssymb,amsfonts}
\usepackage{algorithmic}
\usepackage{graphicx}
\usepackage{textcomp}
\usepackage{xcolor}
\usepackage{comment}
\usepackage[utf8]{inputenc}
\usepackage{textcomp}
\usepackage{xcolor}
\usepackage{tabularx}
\usepackage[hidelinks]{hyperref}
\usepackage{tikz}
\usepackage{todonotes}
\usepackage{circledsteps}
\usepackage{enumitem}
\usepackage{tablefootnote}
\usepackage{threeparttable}
\usepackage{booktabs}
\usepackage{threeparttablex}
\usepackage[cjk]{kotex}
\usepackage{enumerate, float, pgf, circuitikz, multirow, longtable, pdfpages, hyperref, array,siunitx, diagbox}
\usepackage{calc}
\usepackage{pgfplots,pgfplotstable}
\usepackage{xtab,booktabs}
\usepackage{lipsum}
\usepackage{makecell}
\usepackage{pifont}
\usepackage{balance}
\usepackage{import}
\usepackage{comment}
\usepackage{subcaption}
\usepackage{authblk}
\usepackage{balance}

\def\BibTeX{{\rm B\kern-.05em{\sc i\kern-.025em b}\kern-.08em
    T\kern-.1667em\lower.7ex\hbox{E}\kern-.125emX}}

\newcommand{\ie}{\textit{i}.\textit{e}. }
\newcommand{\eg}{\textit{e}.\textit{g}. }
\newcommand{\projectname}{LightEMU}

\usepackage{xcolor}
\usepackage{textcomp}
\usepackage{listings}
\usepackage{lstfiracode}

\usepackage{armasm}
\definecolor{maroon}{cmyk}{0, 0.87, 0.68, 0.32}
\definecolor{halfgray}{gray}{0.55}
\definecolor{ipython_frame}{RGB}{207, 207, 207}
\definecolor{ipython_bg}{RGB}{247, 247, 247}
\definecolor{ipython_red}{RGB}{186, 33, 33}
\definecolor{ipython_green}{RGB}{0, 128, 0}
\definecolor{ipython_cyan}{RGB}{64, 128, 128}
\definecolor{ipython_purple}{RGB}{170, 34, 255}

\definecolor{comment}{rgb}{0,.6,0}
\definecolor{keyword}{rgb}{.63,0,.42}
\definecolor{kw2}{rgb}{.50,.50,.15}
\definecolor{kw3}{rgb}{.42,.42,.63}
\definecolor{string}{rgb}{1,0,0}

\lstdefinestyle{cfira}{
    language=C,
    style=FiraCodeStyle,
    basicstyle=\footnotesize\ttfamily,
}

\lstdefinestyle{c}{
    language=C,
    style=FiraCodeStyle,
    commentstyle=\color{comment},
    stringstyle=\color{string},
    keywordstyle=\bfseries\color{green!40!black},
    identifierstyle=\color{blue},
    basicstyle=\fontfamily{SourceCodePro-TLF}\scriptsize,
    numbers=left,
    numberstyle=\tiny,
    numbersep=5pt,
    frame=lines,
    breaklines=true,
    prebreak=\raisebox{0ex}[0ex][0ex]{\ensuremath{\hookleftarrow}},
    showstringspaces=false,
    upquote=true,
    tabsize=2,
    xleftmargin=2em,
    xrightmargin=2em,
    morekeywords={TEE_Param, TEE_Result}
}

\lstdefinestyle{arma}{
    language=[armasm]Assembler,
    commentstyle=\color{comment},
    stringstyle=\color{string},
    keywordstyle=[1]\bfseries\color{keyword},
    keywordstyle=[2]\bfseries\color{kw2},
    keywordstyle=[3]\bfseries\color{kw3},
    basicstyle=\footnotesize\ttfamily,
    numbers=left,
    numberstyle=\tiny,
    numbersep=5pt,
    frame=lines,
    breaklines=true,
    prebreak=\raisebox{0ex}[0ex][0ex]{\ensuremath{\hookleftarrow}},
    showstringspaces=false,
    upquote=true,
    tabsize=4,
    xleftmargin=2em,
    xrightmargin=2em,
    escapechar=@,
}

\lstdefinelanguage{iPython}{
morekeywords={access,and,break,class,continue,def,del,elif,else,except,exec,finally,for,from,global,if,import,in,is,lambda,not,or,pass,print,raise,return,try,while},%
morekeywords=[2]{abs,all,any,basestring,bin,bool,bytearray,callable,chr,classmethod,cmp,compile,complex,delattr,dict,dir,divmod,enumerate,eval,execfile,file,filter,float,format,frozenset,getattr,globals,hasattr,hash,help,hex,id,input,int,isinstance,issubclass,iter,len,list,locals,long,map,max,memoryview,min,next,object,oct,open,ord,pow,property,range,raw_input,reduce,reload,repr,reversed,round,set,setattr,slice,sorted,staticmethod,str,sum,super,tuple,type,unichr,unicode,vars,xrange,zip,apply,buffer,coerce,intern},%
sensitive=true,%
morecomment=[l]\#,%
morestring=[b]',%
morestring=[b]",%
morestring=[s]{'''}{'''},% used for documentation text (mulitiline strings)
morestring=[s]{"""}{"""},% added by Philipp Matthias Hahn
morestring=[s]{r'}{'},% `raw' strings
morestring=[s]{r"}{"},%
morestring=[s]{r'''}{'''},%
morestring=[s]{r"""}{"""},%
morestring=[s]{u'}{'},% unicode strings
morestring=[s]{u"}{"},%
morestring=[s]{u'''}{'''},%
morestring=[s]{u"""}{"""},%
literate=
    {á}{{\'a}}1 {é}{{\'e}}1 {í}{{\'i}}1 {ó}{{\'o}}1 {ú}{{\'u}}1
{Á}{{\'A}}1 {É}{{\'E}}1 {Í}{{\'I}}1 {Ó}{{\'O}}1 {Ú}{{\'U}}1
{à}{{\`a}}1 {è}{{\`e}}1 {ì}{{\`i}}1 {ò}{{\`o}}1 {ù}{{\`u}}1
{À}{{\`A}}1 {È}{{\'E}}1 {Ì}{{\`I}}1 {Ò}{{\`O}}1 {Ù}{{\`U}}1
{ä}{{\"a}}1 {ë}{{\"e}}1 {ï}{{\"i}}1 {ö}{{\"o}}1 {ü}{{\"u}}1
{Ä}{{\"A}}1 {Ë}{{\"E}}1 {Ï}{{\"I}}1 {Ö}{{\"O}}1 {Ü}{{\"U}}1
{â}{{\^a}}1 {ê}{{\^e}}1 {î}{{\^i}}1 {ô}{{\^o}}1 {û}{{\^u}}1
{Â}{{\^A}}1 {Ê}{{\^E}}1 {Î}{{\^I}}1 {Ô}{{\^O}}1 {Û}{{\^U}}1
{œ}{{\oe}}1 {Œ}{{\OE}}1 {æ}{{\ae}}1 {Æ}{{\AE}}1 {ß}{{\ss}}1
{ç}{{\c c}}1 {Ç}{{\c C}}1 {ø}{{\o}}1 {å}{{\r a}}1 {Å}{{\r A}}1
{€}{{\EUR}}1 {£}{{\pounds}}1
{^}{{{\color{ipython_purple}\^{}}}}1
{=}{{{\color{ipython_purple}=}}}1
{+}{{{\color{ipython_purple}+}}}1
{*}{{{\color{ipython_purple}$^\ast$}}}1
{/}{{{\color{ipython_purple}/}}}1
{+=}{{{+=}}}1
{-=}{{{-=}}}1
{*=}{{{$^\ast$=}}}1
{/=}{{{/=}}}1,
literate=
    *{-}{{{\color{ipython_purple}-}}}1
{?}{{{\color{ipython_purple}?}}}1,
identifierstyle=\color{black}\ttfamily,
commentstyle=\color{ipython_cyan}\ttfamily,
stringstyle=\color{ipython_red}\ttfamily,    % keepspaces=true,
basicstyle=\footnotesize\ttfamily,
numbers=left,
numberstyle=\tiny,
numbersep=5pt,
frame=lines,
breaklines=true,
prebreak=\raisebox{0ex}[0ex][0ex]{\ensuremath{\hookleftarrow}},
showstringspaces=false,
upquote=true,
tabsize=4,
xleftmargin=2em,
xrightmargin=2em,
escapechar=@,
keywordstyle=\color{ipython_green}\ttfamily,
}

\title{\projectname: Hardware Assisted Fuzzing of Trusted Applications}
\author[1]{Haoqi Shan}
\author[5]{Sravani Nissankararao}
\author[2]{Yujia Liu}
\author[3]{Moyao Huang}
\author[3]{Shuo Wang}
\author[4]{Yier Jin}
\author[5]{Dean Sullivan}
\affil[1]{CertiK, \textit{haoqi.shan@certik.com}}
\affil[2]{Li Auto Inc., \textit{liuyujia1@lixiang.com}}
\affil[3]{University of Florida, \textit{moyaohuang@ufl.edu, shuo.wang@ece.ufl.edu}}
\affil[4]{University of Science and Technology of China, \textit{jinyier@ustc.edu.cn}}
\affil[5]{University of New Hampshire, \textit{\{sravani.nissankararao, dean.sullivan\}@unh.edu}}

\begin{document}

\maketitle

\begin{abstract}
Trusted Execution Environments (TEEs) are deployed in many CPU designs because
of the confidentiality and integrity guarantees they provide. ARM TrustZone is a
TEE extensively deployed on smart phones, IoT devices, and notebooks.
Specifically, TrustZone is used to separate code execution and data into two
worlds, normal world and secure world. However, this separation inherently
prevents traditional fuzzing approaches which rely upon coverage-guided feedback
and existing fuzzing research is, therefore, extremely limited. In this paper,
we present a native and generic method to perform efficient and scalable
feedback-driven fuzzing on Trusted Applications (TAs) using ARM CoreSight. We
propose LightEMU, a novel fuzzing framework that allows us to fuzz TAs by
decoupling them from relied TEE. We argue that LightEMU is a promising
first-stage approach for rapidly discovering TA vulnerabilities prior to
investing effort in whole system TEE evaluation precisely because the majority
of publicly disclosed TrustZone bugs reside in the TA code itself. We implement
LightEMU and adapt it to Teegris, Trusty, OP-TEE and QSEE and evaluate 8 real-world TAs while triggering 3 unique crashes and achieving x10 time speedup when
fuzzing TAs using the state-of-the-art TrustZone fuzzing framework.
\end{abstract}


\section{Introduction}\label{noemu:sec:intro}

Fuzz testing ARM TrustZone~\cite{trustzone} software is imperative, especially
due to the significant presence of devices in the market that depend on Trusted
Execution Environments (TEEs) to ensure their security. Indeed, ARM has shipped
nearly 200 billion chips and is the most widely used instruction
set~\cite{chips2021arm}. This can, in part, be attributed to ARM TrustZone's
ability to provide strong confidentiality and integrity guarantees. As such,
many critical applications rely on TrustZone to securely provision and deploy
essential resources. Despite its extensive usage, however, there are
surprisingly few works investigating TrustZone security using automated methods. 

A review of related work revealed only three existing systematic
efforts~\cite{partemu_2020, teezz_2023, shan2023crowbar} to fuzz TEEs. This is
topically intuitive because TEEs are built upon strong hardware-backed isolation
guarantees that naturally prevent fuzzing. Moreover, major vendors lock down
their TEE implementations before market release and share limited implementation
information. ARM's business model further complicates matters with its diverse
inventory of soft processor core IP licensed to vendors, resulting in a variety
of unique products~\cite{arm_ip} with many configurations and features that make
it difficult to provide a fuzzing solution.

\begin{figure*}[!hbtp]
  \centering
  \includegraphics[width=0.95\textwidth]{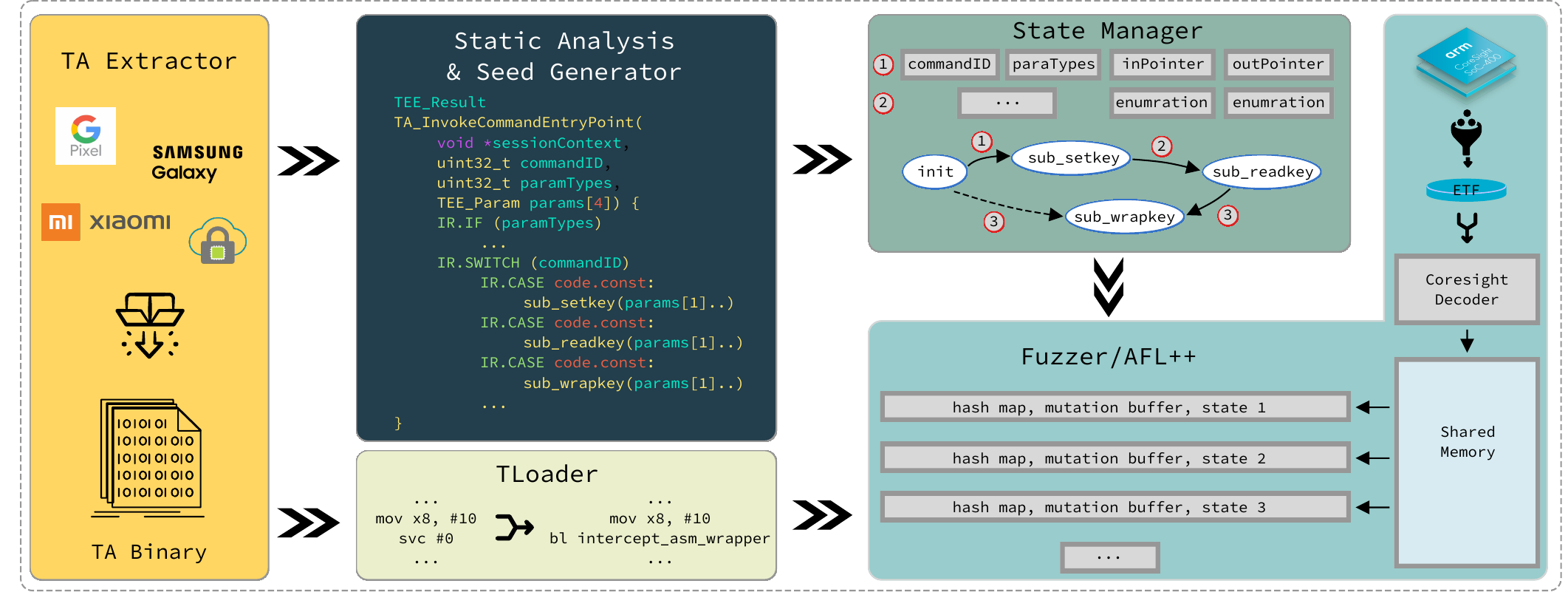}
  \caption{\projectname{} Architecture Design}
  \label{noemu:fig:impl}
  \vspace{-0.5cm}
\end{figure*}

In this paper, we fill this gap with \projectname{}, a generic fuzzing solution
that can be used by developers and researchers alike to fuzz TrustZone
Applications (TAs). The insight underlying \projectname{} stems from an in-depth
analysis of extant bugs affecting TEEs, in which we find that a significant
number of reported Common Vulnerabilities and Exposures (CVEs) in TAs are
independent of the underlying TrustZone hardware, peripherals, or OS (TZOS). As a
consequence, we reason that it is not only possible but also effective, to fuzz
TAs by lifting them from their device and TZOS dependencies. \projectname{}
allows one to fuzz arbitrary TAs at scale without the need for instrumentation
or full device/TZOS emulation on any CoreSight-equipped ARM Cortex-A platform.
In practice, \projectname{} consists of a TA extractor, a static binary analyzer
capable of recovering command IDs and input templates for crafting appropriate
fuzzing seeds, a custom loader (T-Loader) that resolves both statically compiled
and dynamic dependencies of an arbitrary TA at runtime, a state manager that
collaborates with AFL++~\cite{AFLplusplus-Woot20} to maintain TA state while
fuzzing, and CoreSight-assisted tracing to achieve native binary-only fuzzing on
TAs, as shown in Figure~\ref{noemu:fig:impl}.

\subsection{Motivation}

\projectname{} addresses some significant limitations of prior work, namely,
reliance on blackbox fuzzing, rehosting techniques and scalability. A blackbox
fuzzer is limited in that it has neither information about the underlying code,
nor is it instrumented~\cite{liang2018fuzzing}. This effectively restricts the
fuzzer to randomly probing the program. Thus blackbox fuzzing drastically
underperforms compared to both coverage-guided (grey box) and
symbolically-driven (white box) fuzzers~\cite{metzman2021fuzzbench}. 

Rehosting techniques, on the other hand, address the limitations of blackbox
fuzzing by providing an emulated environment on which the embedded binary and
firmware can execute. Coverage-guided feedback can be provided in rehosted
environments because all of the underlying execution is managed by the emulator,
including its control flow. However, rehosting solutions depend on detailed
knowledge of the underlying firmware and hardware it aims to emulate, which, as
previously mentioned, is entirely lacking for ARM TrustZone. 

TEEzz~\cite{teezz_2023} and PartEMU~\cite{partemu_2020} are two recent
approaches implementing the aforementioned techniques. TEEzz is a blackbox
fuzzer that captures format, parameter, and type information from the TA
application programming interface (API) as the TA executes to reconstruct
fuzzing candidates. Mutation of these candidates guarantees not only that a
known state in the TA will likely be reached, but also ensures guided probing of
the TA code base. Albeit TEEzz addresses various shortcomings of fuzzing TEEs,
it has several impractical prerequisites including access to fully-rooted
devices, availability of multiple CAlibs~\footnote{CALib is a wrapper library to
provide regular applications a universal interface to interact with the Trusted
world.}, and only works on specific TEE platforms. These requirements, along
with the laborious manual effort needed for the specified fuzzing template
extractor, render TEEzz's approach unsuitable for general and efficient use.
PartEMU is a whole system TEE rehosting solution. It is the first, and only, of
its kind largely because it was built in-house using proprietary details from
closed-source commercial TZOSes. While their study was able to uncover several
bugs in TAs across various vendors, it suffered from significant slowdowns due
to emulation and is impractical to recreate. 

Crowbar~\cite{shan2023crowbar} presented an orthogonal approach to fuzzing TAs
by utilizing rare prototype devices to natively fuzz TAs on real devices.
However, the approach is highly limited in that it relies on highly specific device
requirements rendering such a fuzzing approach neither scalable nor generic.

\projectname{}'s design allows one to bypass not only issues of blackbox fuzzing
imposed by TA confidentiality and slowdowns due to rehosting, but also
circumvent impractical constraints and reliance on propriety knowledge required
by state-of-the-art solutions~\cite{partemu_2020, teezz_2023}. \projectname{}
performs coverage-guided fuzzing on TAs without the need for instrumentation or
full device/TZOS emulation on any CoreSight-equipped ARM Cortex-A platform. 
We consider \projectname{} a drop-in approach that solves many existing challenges, but importantly, do not argue that it is a replacement for either TEEzz or PartEMU. Instead, we believe \projectname{} is a supplement to be deployed to rapidly assess the likelihood of finding vulnerabilities in a TA prior to any additional analysis effort because \projectname{} both scales to arbitrary TA availability while achieving near-native execution speeds.

\subsection{Contributions}

In summary, we make the following research contributions:
\begin{itemize}
    \item We conduct a comprehensive investigation on the cause of TrustZone
    vulnerabilities and find that the majority of bugs reside in TAs rather
    than peripherals or TZOS.
    \item We propose \projectname{}, a CoreSight-assisted coverage-driven
    fuzzer that enables us to fuzz TAs without the need for emulation
    or a strict hardware environment. 
    \item We implement \projectname{} and support fuzzing TAs from multiple
    vendors, \ie, Samsung Teegris, Google Trusty, Qualcomm QSEE, and OP-TEE
    within the normal world with native execution speed. We evaluate
    \projectname{} on 8 TAs and found 3 unique crashes. We show \projectname{}
    can achieve x10 times speedup compared to the state-of-the-art TrustZone
    fuzzing frameworks.

\end{itemize}

In what follows, we first outline background in
Section~\ref{noemu:sec:background}. We then present an analysis of TrustZone
vulnerabilities in Section~\ref{noemu:sec:rethink}, which leads to a presentation
of \projectname{} in Section~\ref{noemu:sec:impl}. Our evaluation of
\projectname{} is presented in Section~\ref{noemu:sec:eval}. We then present
related works in Section~\ref{noemu:sec:related} before concluding in
Section~\ref{noemu:sec:conclusions}.

\section{Background} \label{noemu:sec:background}
    \noindent In this section, we present the relevant background on ARM TrustZone architecture and ARM CoreSight architecture.
\subsection{ARM TrustZone architecture}

\begin{figure}[htbp]
    \centering
    \includegraphics[width=\columnwidth]{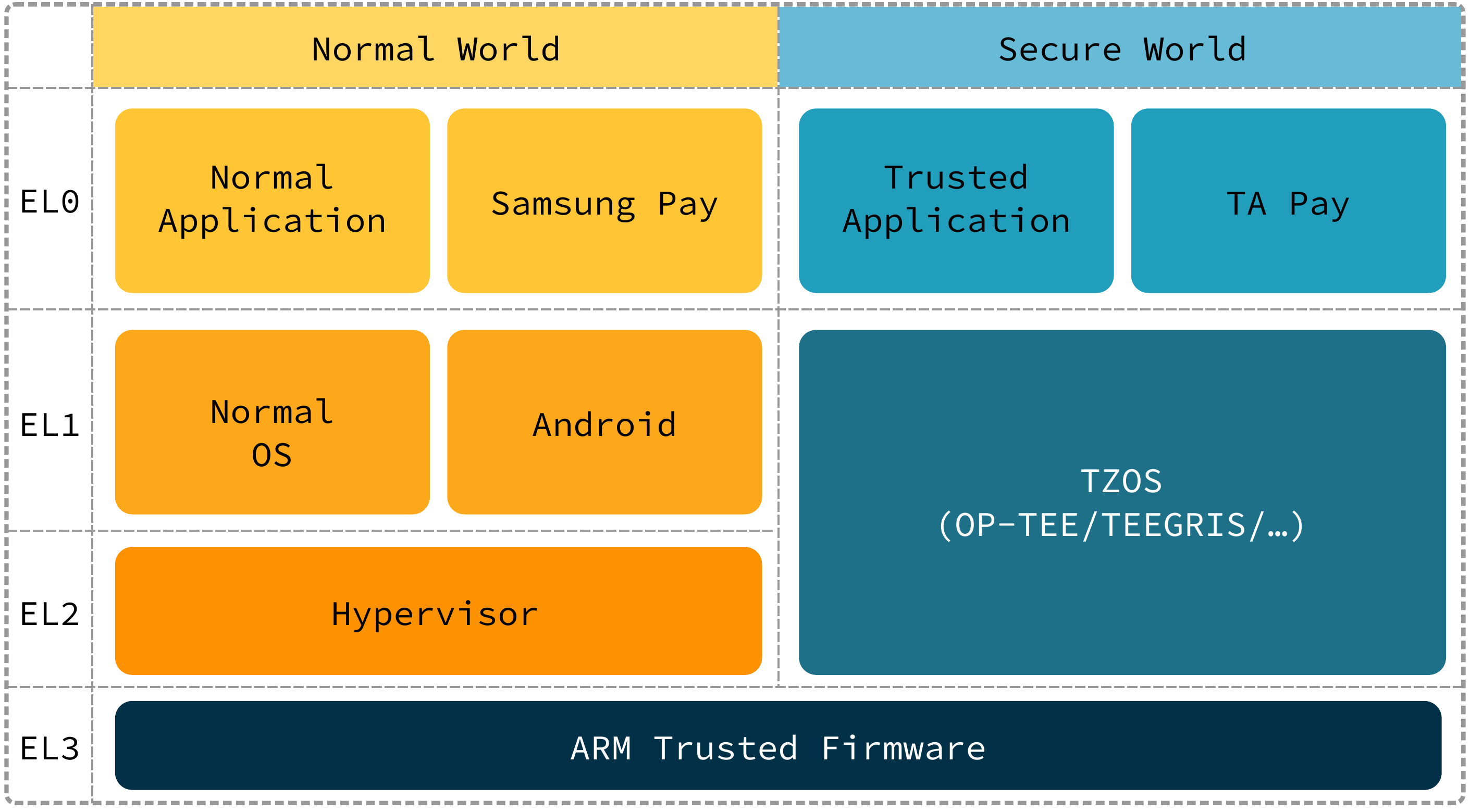}
    \caption{ARM TrustZone architecture.}
    \label{lightemu:fig:trustzone_arch} 
\end{figure}

TrustZone is a security extension technology embedded in ARM processors. It
provides a way to partition a system into two distinct execution environments:
the normal world or Rich Execution Environment (REE) and the secure world or
Trusted Execution Environment (TEE), as shown in Figure~\ref{lightemu:fig:trustzone_arch}. The REE is where the general software
stacks like Linux and Android operate. It is the environment in which most
applications and operating system components run. The TEE is a highly secure and
isolated environment responsible for hosting security-critical applications like
cryptographic services, digital wallets, and digital rights management (DRM). Each world possesses
dedicated memory to execute its respective operating systems and
services. However, the secure world holds higher access privileges, allowing it
to utilize resources from the normal world, including memory and
peripherals, but not vice versa. Additionally, ARM introduces a novel processor
mode, referred to as the monitor mode, which facilitates the transition between
two worlds. The Non-Secure (NS) bit in the secure configuration register (SCR)
indicates the current world in which the processor is executing. When the NS bit
is set, it indicates that the processor is functioning within the normal world,
conversely, when it is not set, the processor is operating in the secure world.

During the world switching, the processor enters the monitor mode through a
Secure Monitor Call (SMC) or an interrupt, in response to a normal world's
requirement. This is handled by ARM Trusted Firmware (ATF)~\cite{atf} and
prevents unauthorized world switch requests. Subsequently, the code
executing in the monitor mode preserves the context of the normal world and
alters the value of the NS bit, thus facilitating the switch to the
secure world. Similarly, the secure world follows the same approach to switch to
the normal world. When Client Applications (CA) in the normal world necessitate
secure services, the processor switches to the secure world, performs the
necessary security operation, and conveys the results back.
\subsection{ARM CoreSight Infrastructure}

\begin{figure}[htbp]
    \centering
    \includegraphics[width=0.9\columnwidth]{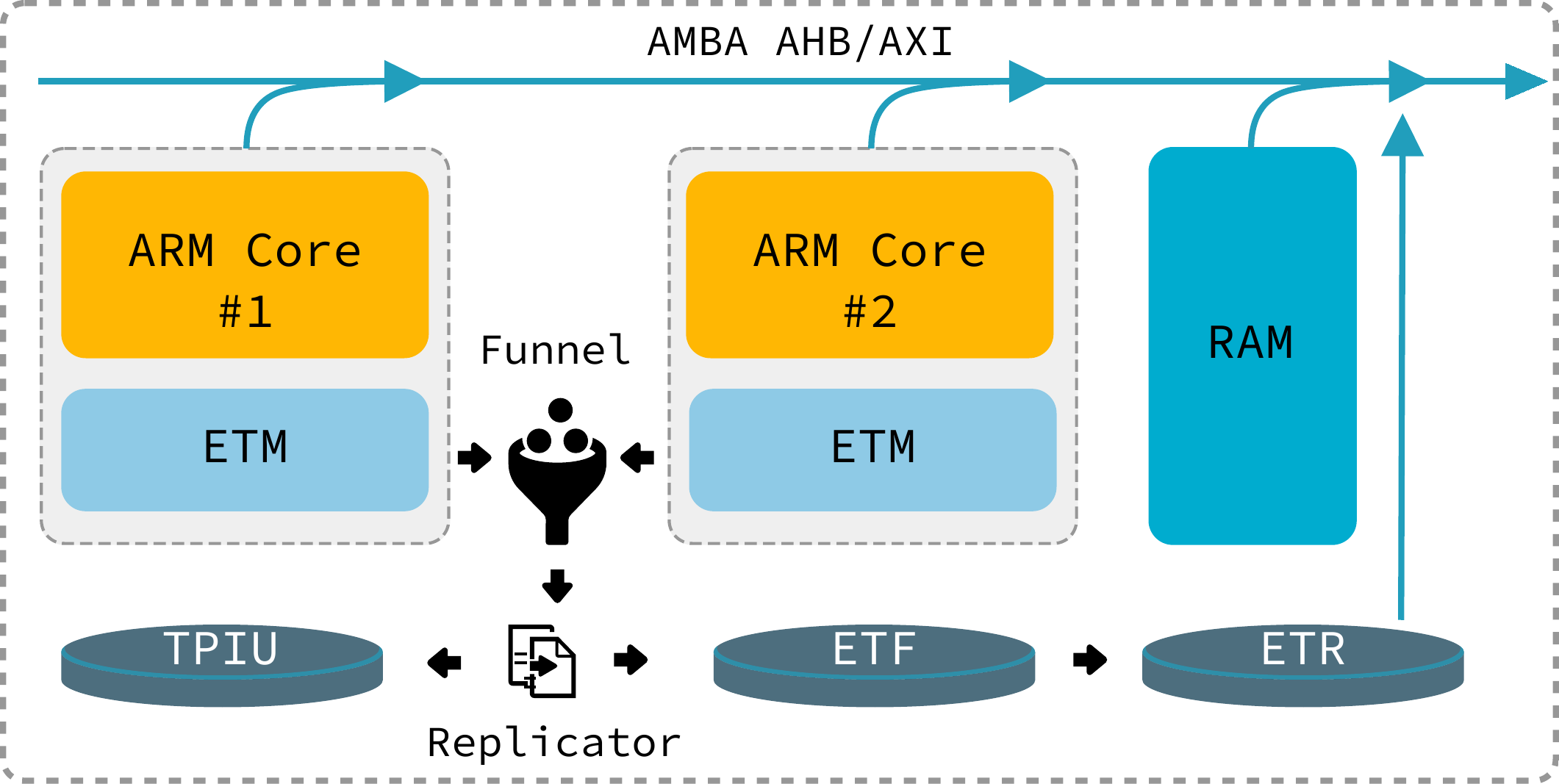}
    \caption{ARM CoreSight architecture.}
    \label{lightemu:fig:coresight_arch} 
\end{figure}

ARM CoreSight~\cite{coresight} is a technology and suite of hardware components
designed to aid in debugging and performance optimization of ARM-based
systems-on-chip (SoCs) and embedded devices. CoreSight offers a range of
features to assist in the development, testing, and analysis of software running
on ARM processors. It establishes a standardized collection of interfaces and
elements that simplify the integration of debugging and tracing functionalities
into ARM processors. A central feature of ARM CoreSight is the introduction of a
dedicated bus shared among all debug trace sources, enabling trace generation
without interrupting the ongoing program execution. The technology, as
illustrated in Figure~\ref{lightemu:fig:coresight_arch}, introduces specialized
units such as the Embedded Trace Macrocell (ETM), which offer versatile
debugging options for processors. To consolidate the generated traces into a
unified stream, the Funnel and the Trace Port Interface Unit (TPIU) are
employed, resulting in integrated trace output. This integrated approach
enhances the ability to monitor and analyze system behavior during debugging and
performance optimization.

\section{Rethinking TrustZone Vulnerabilities} \label{noemu:sec:rethink}
    
\newcommand{\cve}[1]{\href{https://www.cvedetails.com/cve/CVE-#1}{\textcolor{blue}{\texttt{\footnotesize #1}}}}
\setlength{\unitlength}{1mm}
\newcommand{\Newdot}{{\leavevmode\put(0,.63){\circle*{0.9em}}}}
\newcommand{\Newcircle}{\leavevmode\put(0,.63){\circle{0.9em}}}

\begin{table*}[t]
    \centering
    \begin{xtabular}{p{4em}|p{4em}|p{14cm}}
        \hfil \textbf{Vendor} & \hfil \textbf{Location} & \hfil \textbf{TA} \\
        \hline\hline
        \multirow{15}{*}{\rotatebox[origin=c]{90}{\textbf{Samsung}}}
        & \rule{0pt}{3ex} \Newcircle{} &
        \cve{2022-33718}, \cve{2022-30717}, \cve{2022-23998}, \cve{2022-22269}, \cve{2022-22266}, \cve{2022-22263}, \cve{2021-25472}, \cve{2021-25453}, \cve{2021-25430}, \cve{2021-25429}, \cve{2021-25406}, \cve{2021-25397}, \cve{2016-11028}, \cve{2018-21051}, \cve{2020-11600}, \cve{2019-20586} \\  [2.5ex]
        \cline{2-3}
        & \rule{0pt}{3ex} \Newdot{} & \cve{2022-30716}, \cve{2022-25824}, \cve{2022-24930}, \cve{2022-24923}, \cve{2022-23997}, \cve{2022-23996}, \cve{2022-23995}, \cve{2022-23994}, \cve{2022-22292}, \cve{2021-25521}, \cve{2021-25520}, \cve{2021-25501}, \cve{2021-25482}, \cve{2021-25447}, \cve{2021-25446}, \cve{2021-25445}, \cve{2021-25441}, \cve{2021-25440}, \cve{2021-25439}, \cve{2021-25438}, \cve{2021-25433}, \cve{2021-25432}, \cve{2021-25431}, \cve{2021-25428}, \cve{2021-25426}, \cve{2021-25418}, \cve{2021-25412}, \cve{2021-25405}, \cve{2021-25334}, \cve{2017-18651}, \cve{2020-13835}, \cve{2020-13834}, \cve{2020-13832}, \cve{2020-12752}, \cve{2020-11604}, \cve{2020-11603}, \cve{2019-20589}, \cve{2019-20588}, \cve{2019-20587}, \cve{2019-20586}, \cve{2019-20585}, \cve{2019-20584}, \cve{2019-20583}, \cve{2019-20571}, \cve{2019-20563}, \cve{2019-20562}, \cve{2019-20560}, \cve{2020-10837}, \cve{2019-20545}, \cve{2019-20537}, \cve{2022-22271}, \cve{2021-25469}, \cve{2021-25468}, \cve{2018-21076}, \cve{2018-21074}, \cve{2018-21067}, \cve{2018-21066}, \cve{2018-21052}, \cve{2018-21050}, \cve{2018-21049}, \cve{2018-21044}, \cve{2020-11604}, \cve{2020-11603}, \cve{2017-18657}, \cve{2017-18656}, \cve{2017-18655}, \cve{2019-20610}, \cve{2019-20607}, \cve{2019-20603}, \cve{2019-20602}, \cve{2019-20596}, \cve{2019-20590}, \cve{2019-20589}, \cve{2019-20588}, \cve{2019-20587}, \cve{2019-20585}, \cve{2019-20584}, \cve{2019-20583}, \cve{2019-20581}, \cve{2019-20571}, \cve{2019-20563}, \cve{2019-20562}, \cve{2019-20560}, \cve{2020-10849}, \cve{2020-10837}, \cve{2020-10836}, \cve{2019-20545}, \cve{2019-20544}, \cve{2019-20537}, \cve{2018-6246}, \cve{2018-5210} \\ [2ex] \hline
        \multirow{5}{*}{\rotatebox[origin=c]{90}{\textbf{Qualcomm}}}
        & \rule{0pt}{3ex} \Newcircle & \cve{2017-18280}, \cve{2016-10423} \\
        [1ex] \cline{2-3}
        & \rule{0pt}{3ex} \Newdot & \cve{2021-1909}, \cve{2021-1923}, \cve{2021-1889}, \cve{2021-1886}, \cve{2019-14130}, \cve{2019-2329}, \cve{2018-5918}, \cve{2017-18300}, \cve{2021-30338}, \cve{2021-30278}, \cve{2021-1894}, \cve{2020-11123}, \cve{2015-6647}, \cve{2015-6639}, \cve{2020-0042}, \cve{2019-14040}, \cve{2018-13906}, \cve{2017-18278}, \cve{2015-9183}, \cve{2015-9174}, \cve{2015-9169}, \cve{2015-9166}, \cve{2015-9162}, \cve{2015-9132}, \cve{2017-9716} \\
        [1ex] \hline
        \multirow{1}{*}{\rotatebox[origin=c]{90}{\textbf{Android}}}
        & \rule{0pt}{3ex} \Newcircle & \cve{2016-10237}, \cve{2020-0403} \\
        [1ex] \cline{2-3}
        & \rule{0pt}{3ex} \Newdot & \cve{2021-34375}, \cve{2016-0825}, \cve{2021-34389}, \cve{2020-0077}, \cve{2020-0076}, \cve{2020-0075}, \cve{2017-6296}, \cve{2017-6295}, \cve{2015-9005}, \cve{2014-9951}, \cve{2014-9949}, \cve{2014-9948}, \cve{2014-9947}, \cve{2015-9003}, \cve{2015-9002}, \cve{2015-9001}, \cve{2015-9000}, \cve{2014-9937}, \cve{2014-9935}, \cve{2014-9932} \\
        [1ex] \hline
        \multirow{2}{*}{\rotatebox[origin=c]{90}{\textbf{HUAWEI}}} &
        \rule{0pt}{3ex} \Newcircle & \cve{2022-41603}, \cve{2022-41602}, \cve{2022-41601}, \cve{2022-41600}, \cve{2022-41599}, \cve{2022-41598}, \cve{2022-41597}, \cve{2022-41595}, \cve{2022-41594}, \cve{2022-41593}, \cve{2022-41592}, \cve{2021-40014} \\
        [1ex] \cline{2-3}
        & \rule{0pt}{3ex} \Newdot{} & \\
        [1ex] \hline
        \multirow{2}{*}{\rotatebox[origin=c]{90}{\textbf{OPTEE}}}
        & \rule{0pt}{3ex} \Newcircle{} & \\
        [1ex] \cline{2-3}
        & \rule{0pt}{3ex} \Newdot{} & \cve{2019-25052}, \cve{2019-1010294}, \cve{2016-6129} \\
        [1ex] \hline
        \multirow{2}{*}{\rotatebox[origin=c]{90}{\textbf{Others}}}
        & \rule{0pt}{3ex} \Newcircle{}& \cve{2020-7958}, \cve{2021-29415} \\
        [1ex] \cline{2-3}
        & \rule{0pt}{3ex} \Newdot{}& \cve{2021-45451}, \cve{2021-45450}, \cve{2020-12929}, \cve{2019-20779}, \cve{2019-20776}, \cve{2017-5648}, \cve{2020-12821} \\
        [1ex] \hline
    \end{xtabular}
    \caption{\includegraphics[width=0.025\textwidth]{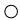} represents peripheral related vulnerability, and \includegraphics[width=0.018\textwidth]{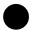} represents non-peripheral related vulnerability}
    \label{noemu:tab:ta_cve}
    \vspace{-0.5cm}
\end{table*}

\projectname{} relies on ARM CoreSight to recover coverage information of TAs. It is natural to then assume that it can be used immediately, \emph{in-situ}, to capture trace information on a TrustZone-protected device. Unfortunately, our analysis revealed several immediate challenges in choosing this direction. To begin with, we found we could not simply naively use the CoreSight IP because it is: 1) unlikely an arbitrary COTS SoC will have licensed it, and even if they had it's unlikely they will have left its interface open once released; and 2) we aim to support general TrustZone fuzzing that is scalable, which is simply not possible when tied to an \emph{in-situ} device.

Ideally, we want to retain CoreSight-assisted fuzzing to overcome the limitation of blackbox fuzzing of TAs without the overhead and scalability issues caused by rehosting. %
We found that we could be confident in doing so because, based on the analysis below, the \textbf{\emph{majority of vulnerabilities reported on TrustZone platforms have not been caused by peripheral or TZOS code, but rather by the TA itself.}}

\subsection{TrustZone Vulnerability Survey Methodology}

The complete listing of CVEs can be found in Table~\ref{noemu:tab:ta_cve}. We summarize our survey results in Figure~\ref{noemu:fig:tas}. 
We use the national vulnerability database of NIST to collect and categorize all publicly disclosed TrustZone vulnerabilities across all vendors. The vulnerabilities in NVD are enumerated with a CVE ID. We found 180 TrustZone-related vulnerabilities by searching the database with the list of keywords: ``trusted application'', ``trustzone", ``QSEE", ``OP-TEE", ``Kinibi", ``Trustonic", ``trustlet'' and ``Teegris". Our keyword list is exhaustive and consequently, we are confident that it is representative. We further describe our categorization as follows.

\smallskip
\noindent\textbf{TA vulnerabilities:} The vulnerabilities occur at the TA level, which is relatively shallow and mostly affects only the TA itself. TA vulnerabilities are therefore categorized based on whether the CVE description refers to: 1) only a specific TA; and 2) a process that fully takes place in userland.

\smallskip
\noindent\textbf{TZOS vulnerabilities:} We classify TZOS vulnerabilities occurring at the kernel level and found they mainly target: 1) The system call interface. For instance, in CVE-2022-40762, the system call function \texttt{TEE\_Realloc} allocates memory with an excessive size which allows any TA that invokes this function to trigger DoS; 2) Privilege escalation. In CVE-2016-2432, the TZOS component allows attackers to gain privileges via a crafted application; and 3) Running process or resource allocation. In CVE-2021-25363, the vulnerability allows an untrusted application to access processes and delete local files.

\smallskip
\noindent\textbf{Peripheral vulnerabilities:} These vulnerabilities require hardware dependencies and are often triggered by TAs that limit access control or enforce Data-Rights Management (DRM) at their exposed interface, such as in a TA fingerprint application
or vulnerabilities that involve hardware like the xPU~\cite{li_qualcomm_nodate}.

\subsection{TrustZone Vulnerability Survey Analysis}

From the 180 vulnerabilities, we identified 146 TA vulnerabilities and 34 TZOS/peripheral vulnerabilities. The survey shows that more than half of vulnerabilities happen within TAs. Furthermore, most of them (81.1\%) do not involve TrustZone peripherals or TZOS code. This indicates, at least topically, that fuzzing the TA by itself will reveal a significant number of vulnerabilities affecting the existing TEE landscape. In what follows, we break down the reported vulnerabilities into vulnerability types to reveal the effectiveness of bug discovery using automated techniques, such as fuzzing.

\begin{figure}[htbp]
    \centering
    \includegraphics[width=\linewidth]{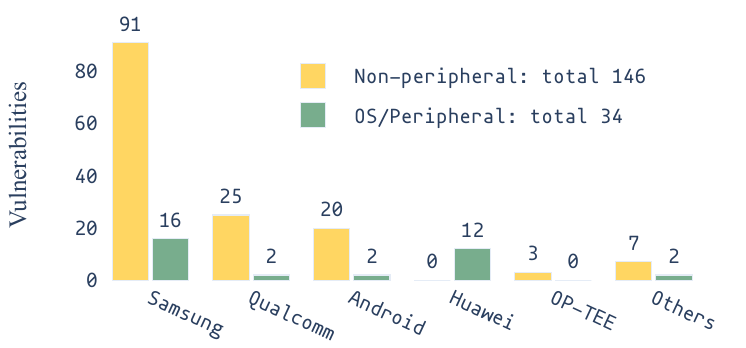}
    \caption{Distribution of vulnerabilities in TrustZone.}
    \label{noemu:fig:tas}
\end{figure}

\begin{figure*}[htbp]
    \centering
    \includegraphics[width=\textwidth]{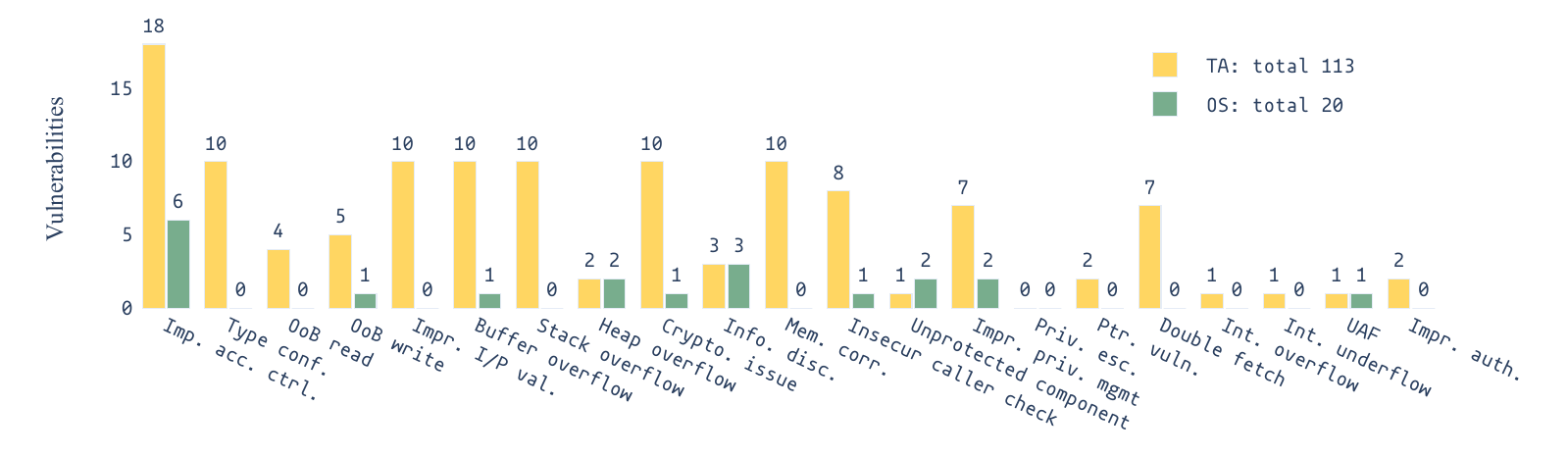}
    \caption{Types of vulnerabilities in TrustZone. }
    \label{noemu:fig:vuln_types}
    \vspace{-0.5cm}
\end{figure*}

Determining the types of bugs that are suitable for fuzzing gives us insight into whether \projectname{} is sufficient to capture the types of vulnerabilities facing TEEs as it relies on TA-only fuzzing. Fuzzers are highly effective at finding bugs related to memory corruption, such as buffer overflows, use-after-free, and other memory-related vulnerabilities. On the other hand, fuzzing is not as effective at discovering deep bugs or issues that require complex setups and interactions with peripheral interfaces and other hardware as is typical in the TZOS. The TZOS is a more complex environment, and its vulnerabilities are likely deeply embedded in the interactions between components, making them harder to trigger using traditional fuzzing methods.

This is reflected in our analysis of vulnerability types shown in Figure ~\ref{noemu:fig:vuln_types}, which shows that of the 133 reported vulnerabilities less than 50\% of TZOS/peripheral vulnerabilities are of the type discoverable with a coverage-guided fuzzer. Note, that we have removed repeated CVE IDs from the raw results reported in both Table~\ref{noemu:tab:ta_cve} and Figure~\ref{noemu:fig:tas} to make our analysis more accurate. Another conclusion we can draw from these results is that 83\% of TA vulnerabilities are well-suited for discovery by fuzzing. The overwhelming majority of reported vulnerabilities relate to memory corruption, which we have broken down into out-of-bounds reads/writes, buffer/stack and heap overflows, information disclosure, and use-after-free depending on the terminology used in the CVE description. In general, we reason that these bugs are common in TAs because they often involve processing untrusted inputs from the normal world, and any mishandling of these inputs can lead to memory corruption issues.

\subsection{An Analysis of TA Complexity}
We are now confident that TA-only fuzzing will provide significant insight into
the security of a TEE-protected device. However, we are still unsure if
coverage-guided fuzzing is best suited for \projectname{}. To address this we
analyzed the TA binary of the High Bandwidth Digital Content Protection (HDCP)
extracted from a Samsung device as a representative sample and compared with
outdated version source code available online. Our analysis includes a breakdown
of the control flow into simple and complex dependencies. Herein, we refer to
anything \texttt{simple} as a control-flow branch that can be easily solved
given the average run-time (\eg 24 hours) of a fuzzer, such as a comparison
against a constant or unknown value. We define a \texttt{complex} dependency
as one whose control flow depends on a pointer to memory (\eg a pointer to a
\texttt{struct} or heap variable). 

\begin{figure}[htbp]
    \centering
    \includegraphics[width=\linewidth]{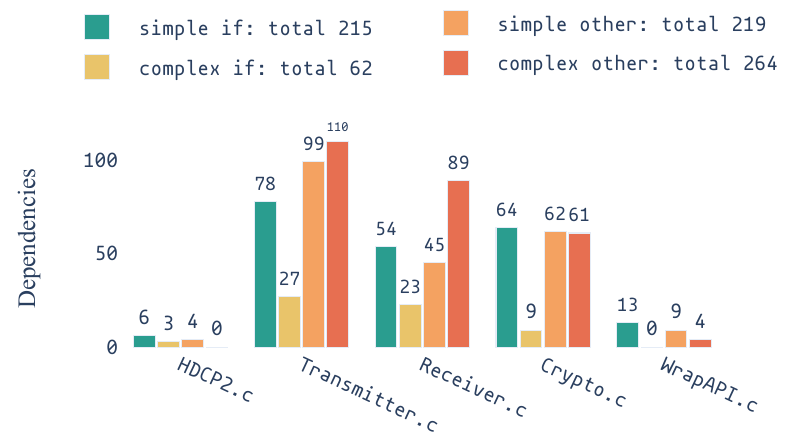}
    \caption{Dependencies breakdown of a sample TA (HDCP).}
    \label{noemu:fig:deps}
    \vspace{-0.5cm}
\end{figure}

From our analysis, shown in Figure~\ref{noemu:fig:deps}, we conclude that 57.1\%
of the TAs code-base are comprised of simple dependencies and 42.9\% of complex
dependencies. In the context of distinguishing between simple and complex
dependencies, manual code review, static analysis tools, and specialized
analysis techniques that leverage domain knowledge are often more effective at
achieving greater code coverage and, therefore, finding more vulnerabilities.
Without these, \projectname{} can likely achieve code-coverage approaching 60\%
of the TA code. It is still probable that a not insignificant
percentage of the remaining code hidden behind complex dependency resolution
will be reached by a traditional coverage-guided fuzzer. However, it is unlikely that mutated inputs will be meaningful enough to allow deep bug analysis
without effort. 

\section{\projectname{} Implementation} \label{noemu:sec:impl}

Herein, we introduce the implementation details of \projectname{} and associated technical challenges on different TZOSes. Designed to address the challenges in TrustZone fuzzing, \projectname{} is built on several key components: a TA extractor for extracting TrustZone Applications from vendor's images, a static binary analyzer for command ID recovery, a custom loader named TLoader for runtime dependency resolution, a state manager that collaborates with AFL++ during fuzzing, and a CoreSight-assisted tracing mechanism for collecting trace during TA executions. This section delves into the intricacies of these components.

\subsection{T-Loader: Normal World Loader for Trusted Applications}

In general, the TAs on an ARM platform have a similar or identical
format to the Executable Linkable Format (ELF)~\cite{wikipedia_elf}. For instance, OP-TEE TAs are
recognized as ELF shared objects by the \texttt{file} command when no encryption is
applied. The similarity between TAs and regular ELF files
creates an opportunity for us to load TAs in the normal world. The reason that we cannot simply use the regular ELF loader to load TAs, such as
\texttt{ld.so}, is that some properties of TAs from certain platforms need to be
parsed ahead of execution, such as TA UUID. %
We, therefore, implement our own loader to load the TAs from different vendors in the normal world. Our implementation is partly inspired by the ELF loader design used by \texttt{OP-TEE} and \texttt{musl}~\cite{musl}.

We compiled, extracted, and analyzed the execution format of TAs from different
vendors and inspected the prerequisites of loading them into memory and executing
them. We found that the loading process of TAs from different vendors varies due to the compilation processes and post-processing steps
on different TZOSes.
For certain
platforms such as OP-TEE and Trusty, the source code is mainly statically
compiled and in general, there is only one TA binary for each TA (although it
does support supplying another TA binary file as a dynamic dependency). For TAs
from Teegris, the TAs are compiled into a shared object and the library
dependencies, such as cryptographic libraries, are dynamically linked and loaded on demand. QSEE is a rather special case, the library for QSEE TAs can be loaded into
a fixed address and the TAs are loaded and linked against this memory region.
Overall, our loader is designed to load both statically and dynamically
compiled TAs into the normal world while adapting all these designs.

\subsubsection{T-Loader implementation}
In our implementation, we parse the TA file and examine the segment
information from its ELF header. We then allocate the memory for parsed segments and
populate them accordingly. During the population, we also iterate through the segments and
locate the dynamic linking information, and resolve dependencies. This
information is pushed into a queue and processed later. We then parse and process
the proprietary information from the TA file if there is any. After that, we
iteratively process and load the dependencies from the queue. The stack size is
also determined in this process and we allocate the stack for TA code before
execution. During the loading process, all memory allocations are designated to
reside in lower memory, such as \texttt{0xff001000} to
\texttt{0xff022000}. These addresses fit into 32-bit integers and are compatible
with 32 bits format TAs as they often expect the memory address pointer to have a size of 32 bits so we can also support 64 bits TAs. 
Once we have the TA loaded into memory, we
save the context of the loader code, populate needed parameters,
and jump to the entrypoint of the TA code along with the correct stack pointer.

\begin{figure}[t]
    \centering
    \includegraphics[width=1.0\linewidth]{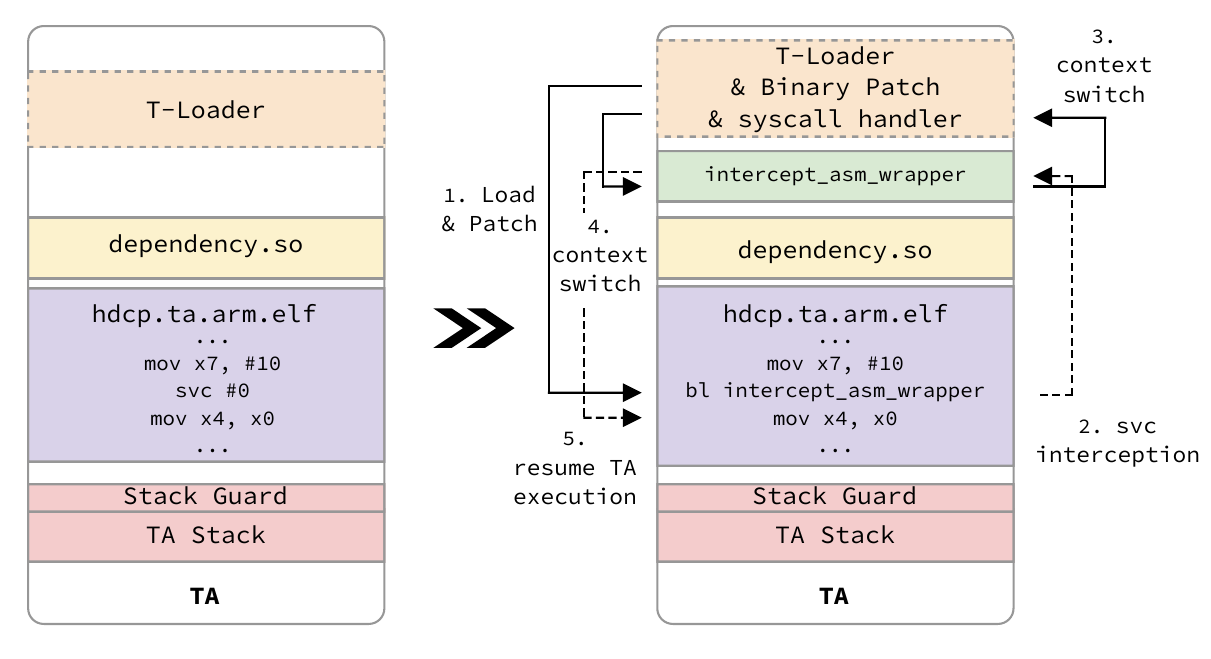}
    \caption{Runtime binary rewrite to hijack \texttt{svc} instructions.}
    \label{noemu:fig:ta_loading}
    \vspace{-0.5cm}
\end{figure}

\subsection{Rewriter: Executing TAs in the Normal World using Dynamic Instrumentation}
However, loading TAs in the normal world is not sufficient for us to actually run them. The TAs only have non-privileged code and they constantly issue system
calls to interact with the TZOS to access hardware or privileged resources. The system
calls used in TAs are different from the system calls used in the regular
normal world. For instance, the syscall numbers are different
and the registers used for holding the syscall number before issuing an \texttt{svc}
instruction are different. To resolve this issue, we implement a dynamic
binary rewriter.

The rewriter engine uses \texttt{capstone}~\cite{capstone} as the disassembler,
parses the TA binary, and locates the \texttt{svc} instruction. We patch the \texttt{svc} instructions with a \texttt{B immi} instruction. The \texttt{B immi} instruction allows us to jump to a \texttt{±32MB} offset from the current
instruction on AARCH64. The offset is dynamically calculated during runtime and
the offset is the address of a trampoline function. The trampoline function
does context switch for us.

For each \texttt{svc} instruction, we have a dedicated trampoline function. After a context switch, the
trampoline function then calls our own \texttt{syscall handler} to handle the system
call and return the result to the TA code. Inside our syscall handler, we check and match the syscall number with our
customized functions to mimic the functionality of syscall handlers of different
TZOSes. Once the syscall handler is completed, we switch back to the TA code and
restore its context. The end of each trampoline function is
dynamically patched to jump back to the next instruction of the corresponding
\texttt{svc} instruction as we need to infer the origin of the \texttt{B immi}
instruction.

\subsection{Tracer: Execution Trace Collection}
    Given that our loader and rewriter now allow us to load and run a TA binary file
in the normal world, one bonus from this achievement is that we can now utilize
the hardware tracing component, CoreSight, to collect the runtime trace of the
TAs without any need for complex binary rewriting or software level overhead
required by emulation. Although the use of T-Loader can be considered as another
way to run TAs in QEMU~\cite{qemu2005} without the need to emulate the entire
TZOS and secure monitor firmware, the main purpose of running TAs natively in
the normal world on ARM platforms that we can achieve no overhead from QEMU
instruction translation using CoreSight's hardware tracing capabilities.

\subsection{Fuzzer: Stateful Fuzzing on Trusted Applications}

Fuzzing TAs poses a significantly different challenge than fuzzing regular
software applications. The main differences are listed as follows: 1) TAs run in
the secure world which makes it difficult for third-party researchers to collect
runtime information; 2) TAs are loaded by the TZOS and will stay in memory until
unloaded. Most importantly, TAs keep a persistent state and wait for the next
invocation from the client even after a previous invocation is completed.
Generally speaking, TAs maintain a state internally and the state of the TAs is
changed with different inputs from the normal world. The stateful design makes
general fuzzing techniques less efficient as it cannot completely traverse the
state machine.

\smallskip
\noindent\textbf{TA Introspection:} As for the first challenge, we have
introduced how to utilize our framework to collect the runtime trace via
CoreSight using T-Loader. Refer to those sections for relevant details and how
we address the technical challenges.

\smallskip
\noindent\textbf{TA Persistence and Statefulness:} The second challenge makes
the biggest difference for \projectname{}. Fuzzing typically proceeds by reading
a mutated seed input. The fuzzed program computes the mutated data and either
finishes execution or exits early. The fuzzer is then able to collect crash
information and determine if the program is vulnerable. However, the stateful
design of TAs makes it impossible to effectively fuzz the TAs with this
traditional fuzzing methodology due to the internal state machine design.

\begin{lstlisting}[
    style=c,
    float,
    floatplacement=H,
    caption={Pseudo code of HDCP TA on Samsung Devices},
    label={noemu:code:hdcp},
    captionpos=b
]

static struct context;

TEE_Result command_dispatch(
    int command, 
    char *request, int request_size, 
    char *response, int *response_size) 
{
    switch (command) {
        case 230:
            return open_crypto_dev(context,...);
        case 231:
            return close_crypto_dev(context,...);
        case 202:
            return init_context(context,...);
        case 222:
            return decrypt_data(context,...);
        case 251:
            return wrap_key(context,...);
        case 252:
            return unwrap_key(context,...);
        case xxx:
            return xxx(...);
        default:
            return TEE_ERROR_INVALID_COMMAND;
    }
}

TEE_Result TA_InvokeCommandEntryPoint(
    void *session, int command, 
    int paramTypes, TEE_Param params[4]) 
{
    if(validate_params(paramTypes, params)) {
        return TEE_ERROR_BAD_PARAMETERS;
    } else {
        // move params into request buffer
        request = ...;
    }

    if(!initialized(command)) {
        return TEE_ERROR_BAD_STATE;
    }

    return command_dispatch(command, request, params[1].buffer, &params[1].size);
}
    \end{lstlisting}

In Listing~\ref{noemu:code:hdcp}, we show the pseudo-code of the HDCP TA on
Samsung devices. The HDCP TA is designed to implement the HDCP
protocol~\cite{hdcp} within TrustZone and export the HDCP-related functions to
the normal world. The HDCP TA involves encrypting/decrypting stream data using
an on-device hardware cryptographic accelerator and associated key management.
The actual HDCP TA that Samsung is currently using supports roughly 50
commands. We reverse-engineered these different commands and their corresponding
command processing functions. As shown in Listing~\ref{noemu:code:hdcp}, the
incomplete pseudo-code includes the entrypoint of the TA, the command processing
function, and static structures to store the context information.

As aforementioned, a CA invokes a TA by sending commands and requests to a
kernel driver. The kernel driver passes the request to the Trusted OS Dispatcher. Eventually, the TA is notified about the request. Irrespective of the extra
operations that are performed by multiple layers of software, the TA and CA are
designed to use a set of predefined parameter conventions, and such parameters
are generally not altered by the OS or ATF~\cite{atf}. These
parameters are passed into the entrypoint of the TA
(\texttt{TA\_CreateEntryPoint} on GP Compliant platforms or
\texttt{CApp\_invoke} on Qualcomm platforms). The entrypoint of the invoke
function has equivalent functionality to the \texttt{main} function in a regular
program and it is invoked every time a new command is received.

The entrypoint functions accept several parameters including the command ID,
request body pointer, request body size, response body pointer, and response
body size pointer. Within the entrypoint function, the TA first checks if these
parameters meet basic requirements, such as NULL pointer check, size check, etc.
Then the TA compares the command ID with the predefined command using, in every
instance we have encountered, a \texttt{switch-case} statement.

For each valid command, the TA executes the corresponding branch function to
process it. When the branch function completes, the TA will return the response
body and response body size to the TZOS and eventually return to the CA. This
process is repeated multiple times until the CA finishes all its requests in the
current session. During these iterations, the CA uses a predefined state machine
to decide which command to send for the next request. The command sequence is
carefully constructed so that the TA can prepare the necessary data for the next
command. As we can see in Listing~\ref{noemu:code:hdcp}, in order to
successfully perform decryption, there are a number of operations needed to be
done first before executing the corresponding command handler, such as
initializing a session context (\texttt{init\_context}), opening the hardware
cryptographic accelerator (\texttt{open\_crypto\_dev}), and unwrapping the key
(\texttt{unwrap\_key}).

\smallskip
\noindent\textbf{Command ID Enumeration:} To effectively fuzz TAs we need to
construct a reasonable parameter set as the input of the fuzzed program. The
parameter set should include a possible command ID that will be processed by the
TA, a valid type of request body, such as either a memory buffer pointer or a
simple integer that can be used as an enumeration value, other parameters, and a
parameter type value to represent the type of these parameters. To collect all
possible command IDs, we utilize the fact that the vast majority of the TAs use
a \texttt{switch-case} statement to process the commands. By leveraging the
binary lifting technique of our T-loader, we can use Binary
Ninja~\cite{binary_ninja} or a similar reverse engineering tool to collect all
switch statements in the TAs and the corresponding constant value that it
compares with.

There are false positive cases that we need to filter out, i.e., a regular
switch case that is declared somewhere other than the request handler. To filter
this out we use taint analysis to identify if the tainted variables that are
used in the switch case are inherited from the entrypoint function. After
filtering out the false positives, the remaining command IDs are collected for
fuzzing. Note, we do not need to find the corresponding CA and
perform binary analysis or symbolic execution to find all used command sequences
in CA code. Instead, we solely focus on the TA to find the command handler
dependencies and use them to reconstruct all state transitions.

\smallskip
\noindent\textbf{Resolve Device Dependency:} As shown in
Listing~\ref{noemu:code:hdcp}, some TAs interact with the hardware via the
device driver. The device driver is usually implemented in kernel space and is
exposed as a userland accessible file, \ie, \texttt{dev://crypto} for the
hardware cryptographic accelerator or \texttt{phys://} for memory access. These
device files are usually opened with \texttt{open} system calls and operated upon
using \texttt{read}, \texttt{write}, and \texttt{ioctl} system calls.

Due to the mechanism of how the TA processes an incoming command, the file
descriptor of these device drivers are often stored as static variables in the
TA's local thread memory. To increase the fuzzing coverage, we identify the use
of such device drivers by first analyzing the TA binary code to find special
files that are called with \texttt{open}. We then filter out the use of these
files by building a call graph of the TA binary that directly uses the file
descriptor variable. Due to the regularity of operations on such file
descriptors, we then identify the command ID that is associated with the
read/write operation to understand the command sequence that is required to use
the device driver.

\smallskip
\noindent\textbf{Resolve Memory Dependency:} Between different commands, the TA
does not just return the processing result to the normal world but instead
stores intermediate results into the secure memory. Such intermediate results
are associated with a session and stored in a context structure. To enhance our
state transition discovery, we also identify the read and write operations for
the context structure. The members and layout of the context structure vary from
TA to TA. The context member is usually not directly used by read/write system
calls, instead, it is typically passed as a pointer to TEE-related functions,
\eg \texttt{TEE\_MemMove}. We use Angr\cite{shoshitaishvili2016state,
stephens2016driller, shoshitaishvili2015firmalice} to perform taint analysis and
trace the memory operation dependency in each individual command handler.
Although there are many TEE functions that can be used to read/write the context
member, eventually these functions will use
\texttt{memcpy}/\texttt{malloc}/\texttt{free} functions to perform memory
operations which give us information about the memory move operations, thereby
allowing us to recover the memory dependency with high fidelity.

\subsection{Case Study}
    \begin{figure}
    \centering
    \includegraphics[width=1\linewidth]{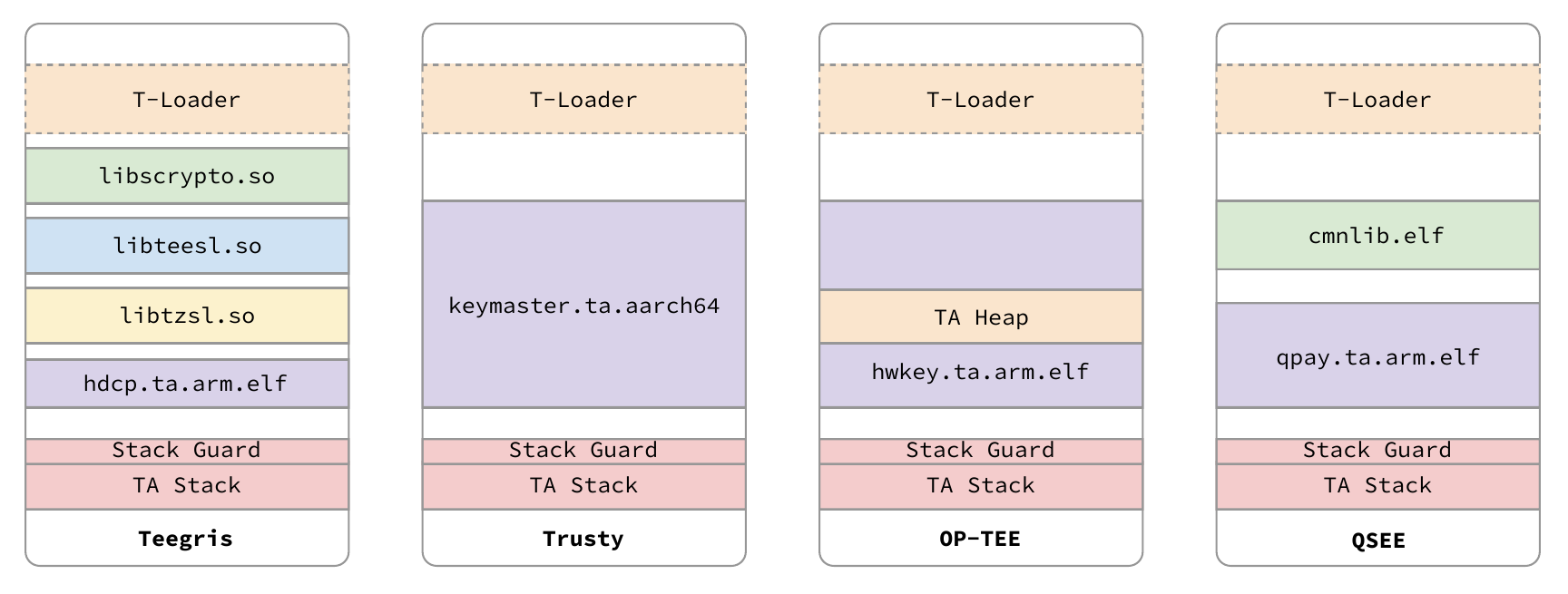}
    \caption{Memory layout of TAs with respect to different TZOS.}
    \label{noemu:fig:loader}
    \vspace{-0.5cm}
\end{figure}

\noindent\textbf{TA Loading Process:} The T-Loader implementations for different
TZOSes vary. The main difference is whether the TA has extra dependencies that
need to be loaded and linked before the TA runs. After reverse engineering, we
find that Trusty and OP-TEE have the TA statically compiled and all the
\texttt{svc} instructions exist in the TA binary. On the other hand, Teegris has
all TAs dynamically compiled and requires the library dependencies
\texttt{libtzsl.so} and \texttt{libscrypto.so} to be loaded ahead of TA
execution. We also find that QSEE TA can be either statically compiled or use
the common library, \texttt{cmblib}, for all the library function calls. The
extra binary dependencies require the T-Loader to serve as dynamic linker and
resolve the external functions at runtime. The memory layout of the loaded TA
from different vendors with T-Loader is illustrated in
Figure~\ref{noemu:fig:loader}.

\smallskip
\noindent\textbf{Syscall Identification:} Another important task of the T-Loader
is redirecting all the syscall functions to our own syscall handler. Within our
syscall trampoline function, we save the context of \texttt{svc} instruction,
such as syscall number, arguments, stack pointer, etc. Then we match the syscall
number with our own syscall table and process the requests. The syscall number
that is used by TZOS is significantly different than the syscall number used by
the Linux kernel although the syscall function name may be reused.

On Teegris, OP-TEE, and Trusty, the syscall number can be directly associated
with the function name that invokes the \texttt{svc} instructions. The functions
usually are just wrappers of the \texttt{svc} and are named with the actual
functionalities. However, on QSEE, such an \texttt{svc} wrapper is implemented
as a generic syscall function and the mapping between the syscall number and the
actual syscall functionalities can only be found via reverse engineering.
Although there are usually more than dozens of syscall functions implemented in
every TZOS, we find that the actual syscall function that is used by TAs is only
a small subset of them. For instance, we analyzed all Samsung TAs for which we
had access, and the syscalls used by them are often \texttt{read},
\texttt{write}, \texttt{close}, \texttt{ioctl}, \texttt{mmap}. As a result, we
only implement the syscall handlers that are actually used by the fuzzed TAs.

\smallskip
\noindent\textbf{Entrypoint and Command Invoke:} To invoke the command handler
of the TAs, we need to customize the callee function that is invoked in the
T-Loader. The command handler function is not necessarily the entrypoint address
defined in the TA ELF header. The implementation of \projectname{} to invoke the
TA on different TZOSes are significantly different and customized for each TZOS.

\smallskip
\noindent\textbf{OP-TEE:} The entrypoint of OP-TEE TA is named as
          \texttt{\_\_ta\_entry} and this function is called with a variable to
          determine if a new session should be created or request should be
          passed with the existing session ID. To invoke the TA command handler,
          the T-Loader first invokes \texttt{\_\_ta\_entry} to create a session
          for initialization purpose. Then for every TA request, the T-Loader
          will setup the arguments and call \texttt{\_\_ta\_entry} again with
          the session ID to invoke the command handler.

\smallskip
\noindent\textbf{Teegris:} The entrypoint of Teegris TA is named
          \texttt{\_start} like a regular Linux ELF file. This function further
          invokes \texttt{\_main} which is implemented in a dependency library
          for initialization. The initialization function sets up the event loop
          that can be used to receive commands from TZOS and invokes the
          initialization function \texttt{TA\_CreateEntryPoint} which is
          implemented in the TA to initialize task-specific resources. The
          requests are dispatched to \texttt{TA\_InvokeCommandEntryPoint}.
          Instead of fully supporting this process, we directly invoke
          \texttt{TA\_CreateEntryPoint} to setup the TA prerequisites and invoke
          \texttt{TA\_OpenSessionEntryPoint}. Then for every new command
          request, we setup the parameters and call
          \texttt{TA\_InvokeCommandEntryPoint} to fuzz the command handler.

\smallskip
\noindent\textbf{QSEE:} Unlike OP-TEE and Teegris, QSEE is not fully GP
          compliant. It means the QSEE TAs are less standardized. The main
          difference in our T-Loader implementation is how to initialize and
          pass the request. The T-Loader does not invoke \texttt{\_main}
          function and expose the TA information to TZOS as QSEE does. Instead,
          we only invoke \texttt{tz\_app\_init} to complete the TA-specific
          initialization. The fuzzing request is passed to the TA by invoking
          \texttt{CApp\_invoke} directly that eventually calls
          \texttt{tz\_app\_cmd\_handler}.

\smallskip
\noindent\textbf{Trusty:} Passing the request to TAs on Trusty is rather
          complicated as Trusty utilizes an IPC design and exposes a set of
          trusted IPC services that can be called. An event loop is created to
          wait for the IPC requests and dispatch them to service handlers,
          similar to Teegris. Instead of going through the event port matching,
          event handler, and message handler as a normal IPC request would we,
          instead, directly identify the message handler and pass the request.
          We believe this is justified because vulnerabilities are unlikely to
          happen in the event dispatching processes, but rather in the message
          handler.

\section{Evaluation} \label{noemu:sec:eval}
    We implemented our fuzzing framework, \projectname{}, on the NVIDIA TX2 dev
board to natively fuzz TAs in the normal world for Teegris, QSEE, Trusty, and
OP-TEE TAs. We ran \projectname{} on 8 TAs that are extracted from firmware
images of Samsung, Xiaomi, and Google and found 3 unique crashes. During the
fuzzing, our fuzzing speed is about 100 requests per second. We further ran
\projectname{} on sample TAs from OP-TEE and recorded all input and CoreSight
traces. We use the same input and run the same trusted application on the NXP
i.MX8M EVK board and record the execution trace within the trusted world. We
compare the two execution traces and show that our \projectname{} trace has high
fidelity to the execution control flow on real hardware.

\subsection{Execution Speed}

We compare the execution speed of \projectname{} with two previous TrustZone
fuzzing approaches, TEEzz~\cite{teezz_2023} and PartEMU~\cite{partemu_2020}.
TEEzz reported the execution speed on fuzzing \texttt{keymaster} and
\texttt{gatekeeper} around 10 requests per second. The execution speed is
relatively the same as our prototype devices' based-fuzzing experiment, which is
\textbf{x10} slower than our \projectname{} implementation. PartEMU did not
report the fuzzing speed, so we performed our own experiment using QEMU for
comparison purposes by running \projectname{} within QEMU. We modified the
fuzzer implementation to gather the execution trace using QEMU instead of
CoreSight. We then load the same TA image that is used in our evaluation and run
the fuzzer. Compared with PartEMU this approach only involves running TAs which
exclude the TZOS or other components. In theory, our experiment should be faster
than PartEMU. We ran the QEMU-based fuzzer for 24 hours and found that although
the fuzzer can trigger the same crashes found while fuzzing on actual ARM
processors, the execution speed is around 8 requests per second which is much
slower than our \projectname{} approach which achieves over 100 requests per
second.

\subsection{Execution Comparison}

\begin{figure}[htbp]
    \vspace{-1em}
    \centering
    \includegraphics[width=\linewidth]{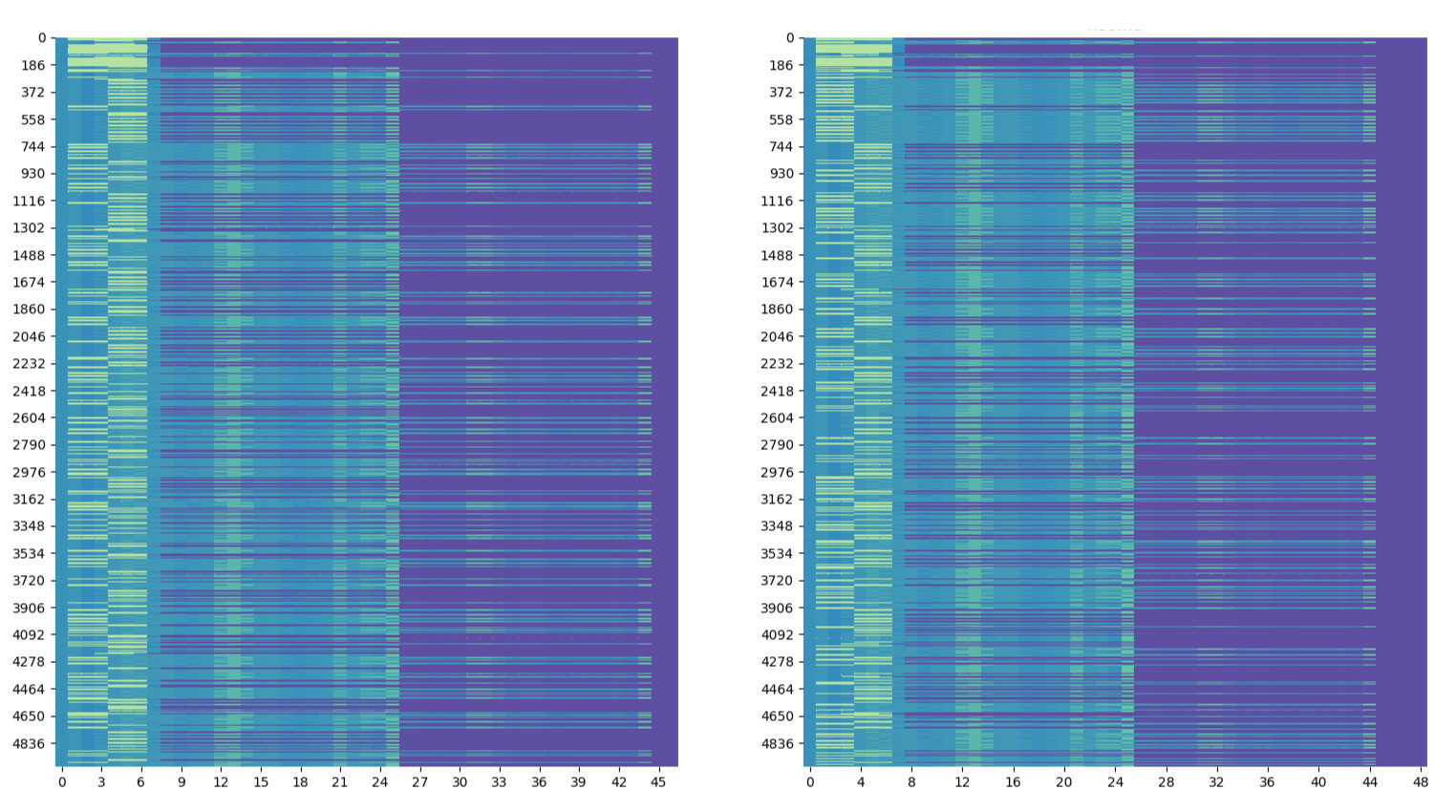}
    \caption{Comparison of TA execution in the secure world (left) and in the normal world (right).}
    \label{noemu:fig:trace_comparison}
    \vspace{-0.3cm}
\end{figure}

It is necessary for us to understand the difference between fuzzing TAs in a
secure world and in a normal world. To do so, we compare the execution traces
of the same TA in the secure world and in the normal world. We designed an example
TA that involves multiple TrustZone operations on the NVIDIA TX2 platform
(Trusty), such as AES encryption/decryption, read/write secure memory region,
read random number generator, etc. The compiled TA is loaded in a secure world
and executed via sending commands and associated ciphertext requests to the TIPC
driver. CoreSight is configured to collect the trace of the TA execution with a
filter set to only capture S-EL0 instructions. The trace is not only used as 
feedback for the fuzzer but also collected to track all the branch
instructions visited by the TA. We also recorded all the input that was sent
into the secure world while fuzzing. We stop the fuzzing after 10,000
iterations. The recorded input is then used to replay the TA execution in the
normal world using \projectname{}. CoreSight is configured to collect the TA
execution trace with a filter set to only capture while the CPU executes the TA
code. The TA execution trace from both the secure world and the normal world is
then normalized with respect to virtual addresses and compared.

We perform full control flow reconstructions with traces from two worlds and map
the visited address into a bitmap. The bitmap is used to illustrate the captured
branch coverage for TA executions, which is similar to our fuzzing feedback
collection process. As shown in Figure~\ref{noemu:fig:trace_comparison}, the
heatmap of the coverage of TA execution within both worlds shows that with the
same input, the TA executions with \projectname{} are similar to the execution
with all the hardware and software dependencies inside the secure world.

\section{Related Work} \label{noemu:sec:related}

Trusted Execution Environment (TEE) fuzzing is a critical aspect of security
research and firmware rehosting through emulation and blackbox fuzzing running
on-device are the two common approaches used to tackle the challenges of
TrustZone fuzzing.

In firmware rehosting, Full rehosting involves creating a detailed emulator that replicates a specific system's behavior, including hardware, using original firmware and extra data for high fidelity and functionality, as seen in Firmadyne~\cite{chen2016towards}, HAL~\cite{HAL_2020}, and Pretender~\cite{gustafson2019toward}. Conversely, Partial rehosting, like P2IM~\cite{feng2020p2im}, focuses only on firmware without additional system or peripheral details. However, both these approaches are suitable for the Rich Execution Environment (REE) rather than the Trusted Execution Environment (TEE).

The challenge of fuzzing TrustZone lies in the intricate balance between
hardware and software dependencies. Harrison et. al's
PartEMU~\cite{partemu_2020} streamlines this process by
providing the essential dependencies for uncovering TEE vulnerabilities. However, this solution poses significant limitations, such as limited fuzzing coverage, coupled with an incomplete simulation
of hardware. Additionally, the
overheads associated with cross-architectural emulation can hinder the fuzzing
process.

Busch et al.'s TEEzz~\cite{teezz_2023} takes a different approach, introducing a
blackbox fuzzing methodology tailored for TAs on COTS devices. By meticulously
analyzing interactions in the normal world, TEEzz discerns the categories and
parameters of TA's APIs. This data is then harnessed to craft specific fuzzing
templates. However, the methodology's dependence on fully unlocked devices,
combined with the manual labor required in defining diverse interface
definitions and its platform specificity, limits its broad applicability.

Shan et al.'s Crowbar~\cite{shan2023crowbar} offers yet another perspective,
focusing on the use of rare prototype devices for the native fuzzing of TAs.
While innovative, this approach's reliance on highly specific device
requirements poses scalability and generic applicability challenges.

In the backdrop of these methodologies, LightEMU stands out with its unique
proposition. Detaching the TA from its TEE enables a more streamlined and
efficient fuzzing of TAs on COTS devices. LightEMU's utilization of ARM
CoreSight ensures that fuzzing is achieved without the typical constraints of
emulation or strict hardware environments. 

\section{Conclusion and Discussion} \label{noemu:sec:conclusions}
    We extensively investigated common causes and dependencies of reported
vulnerabilities in TEEs. This analysis revealed that most CVEs attributed to
TEEs reside in the TA codebase alone. Therefore, we propose \projectname{}, a
solution that decouples the TA from its TEE to rapidly and scalably fuzz them on
COTS devices for vulnerabilities. We 
believe \projectname{} works as a first-stage analyzer for justifying the effort
required to either rehost or black box fuzz whole system TEEs. Our results are
promising and achieve high fidelity compared to native execution. We implement
\projectname{} and adapt it to Teegris, Trusty, OP-TEE, and QSEE and evaluate 8
real-world TAs while triggering 3 unique crashes and achieving x10 time speedup
when fuzzing TAs compared to the start-of-the-art TrustZone fuzzing frameworks.

Although our framework currently only works on the ARM Cortex-A platform, we plan to explore the possibility of extending our hardware-assisted lightweight rehosting fuzzing technique to ARM Cortex-M and Intel SGX platforms in the future.

\section{Acknowledgement} \label{noemu:sec:acknowledgement}
    We appreciate the reviewers for all their constructive suggestions. This work is supported by U.S. Department of Energy under award number DE-SC0018430. 

\balance
\bibliographystyle{IEEEtran}
\bibliography{lightemu}

\end{document}